# Core–shell NaErF$_4$@NaYF$_4$ upconversion nanoparticles qualify a NIR speckle wavemeter by a visible CCD

Tianliang Wang[1], Yi Li[*,1], Long Yan[2], Qin Liang[1], Xu Wang[1], Jinchao Tao[1], Jing Yang[1], Yanqing Qiu[1], Yanlong Meng[1], Bangning Mao[1], Shilong Zhao[1], Pengwei Zhou[1], Bo Zhou[*,2]

1. *College of Optical and Electronic Technology, China Jiliang University, Hangzhou 310018, China*
2. *State Key Laboratory of Luminescent Materials and Devices, Guangdong Provincial Key Laboratory of Fiber Laser Materials and Applied Techniques, and Guangdong Engineering Technology Research and Development Center of Special Optical Fiber Materials and Devices, South China University of Technology, Guangzhou 510641, China.*

E-mail: yli@cjlu.edu.cn; zhoubo@scut.edu.cn

## Abstract

Speckle patterns have been widely confirmed that can be utilized to reconstruct the wavelength information. In order to achieve higher resolution, a varies of optical diffusing waveguides have been investigated with a focus on their wavelength sensitivity. However, it has been a challenge to reach the balance among cost, volumes, resolution, and stability. In this work, we designed a compact cylindrical random scattering waveguide (CRSW) as the light diffuser only by mixing TiO$_2$ particles and ultra-violate adhesive. The speckle patterns are generated by the light multiple scattering in the CRSW. Importantly, the thin layer of upconversion nanoparticles (UCNPs) were sprayed on the end face of the CRSW. This allows the near-infrared (NIR) light to be converted to the visible light, breaking the imaging limitation of visible cameras in the NIR range. We further designed a convolution neural network (CNN) to recognize the wavelength of the speckle patterns with good robustness and excellent ability of transfer learning, resulting in the achievement of a high resolution of 20 kHz (~ 0.16 fm) at around 1550 nm with temperature resistance of ±2°C. Our results provide a low-cost, compact, and simple NIR wavemeter in particular with the ultra-high resolution and good temperature stability.

# TOC Graphic

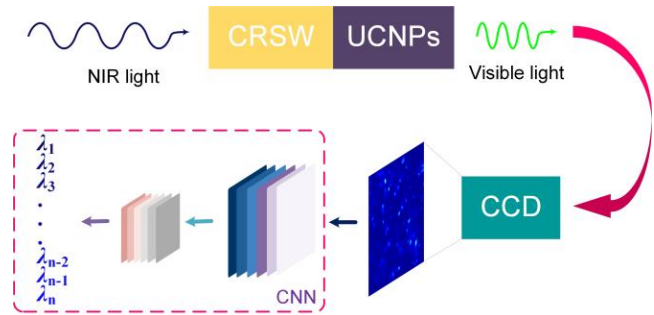

# Introduction

Optical wavemeters play a crucial role in many research fields. In the area of atomic physics, the wavemeter is utilized for probing the interactions of thermal atoms and feedbacking the clock laser wavelength of the atomic clock. [1, 2] While for optical sensing, it can be used for the correction of nonlinear frequency scan in broad applications of optical frequency domain reflectometry (OFDR), optical coherence tomography (OCT), and frequency modulated continuous wave (FMCW) LiDAR [3-5]. Additionally, in optical communications, wavemeters can be used as the equipment of coherence optical spectrum analyzing and the channel monitoring [6, 7]. These applications are raising high requirements for the performance of wavemeter, such as resolution, volume, cost, and speed of measurement. The traditional wavemeters rely heavily on interferometers such as Michelson, Fizeau, and Fabry–Perot [8-10], however, they cannot satisfy the requirements of multi-scenario applications because of their large volume, complex structure and associated high-cost.

By contrast, the laser speckle based wavemeter provides a better solution to overcome these challenges [11, 12]. When a coherent light incident propagates into a multi-mode waveguides or disordered media, a lot of granular speckle patterns can be easily captured by a camera. Goodman first pointed out that the speckles are frequency dependent [13]. Based on this intrinsic characteristic, the speckle patterns have been used to measure laser wavelength and spectrum. Recent researches have been focused on the speckle patterns with stronger wavelength sensitivity for higher resolution wavelength and spectrum measurements. In the beginning, the multimode fiber (MMF) was firstly used as the laser speckle generator [14]. Extending the length of the MMF is the most direct method to increase the resolution[15, 16]. However, longer MMF will bring new defect because it is very susceptible to external environmental influences (especially temperature). This may lead to random changes in the speckle patterns that are not caused by the wavelength variations. In addition, it is quite difficult to keep the temperature invariant when MMF is tens of meters long. These disadvantages seriously compromise the stability of the wavemeter. In the following researches, solutions such as integrating sphere, tapered waveguides, and speckle external modulation have successfully improved the resolution, but they still cannot achieve the balance between volumes, resolution, and stability [17-21]. Another limitation of current speckle wavemeters to wider applications is their cost, especially in the NIR range. In order to capture more details of speckle patterns, InGaAs cameras were indispensable in the measurement of NIR lasers. However, InGaAs cameras are extremely expensive, simultaneously with shortcomings of low-resolution, pixel defects, and the bulky volume [14, 15]. By contrast, the silicon (Si) based cameras are of much lower cost, higher resolution, and easy to drive. If the Si camera can be used in the NIR band, the above problems will be resolved completely.

In this work, we proposed and demonstrated a high resolution, simple, low cost and stable NIR wavemeter by combining the upconversion nanoparticles (UCNPs) and Si-based camera. The pivotal components of this design is a cylindrical random scattering waveguide (CRSW) with a thin layer of UCNPs at the output end. As the diffusing media, the CRSW is composed of random distributed titanium dioxide ($TiO_2$) micro particles embedded in the optical adhesives (OA). The multiply scattering is easier to gain longer optical path than other multi-mode waveguides. More importantly, the concept of photon upconversion provides possibilities to utilize the NIR light especially at around 1530 and 1550 nm. We recently demonstrated that $NaErF_4@NaYF_4$ core-shell UCNPs are good candidate to achieve broad photon upconversion under the NIR laser irradiation [22, 23]. By spraying these core-shell UCNPs on the end face of the CRSW, the NIR light can be converted to visible light after multiple scattering. Then, these visible patterns can be captured by a low-cost and high-resolution visible camera. In order to eliminate the impact from the ambient light and shot noise, we designed a convolution neural network (CNN) for the wavelength reconstruction. The experimental results show that the resolution of the wavemeter achieves to 30 kHz with a broader measurement range.

## Experimental Results

### Experimental setup

The proposed wavemeter is consisted of a piece of polarization maintaining fiber (PMF), a CRSW with a thick layer of UCNPs, and a visible monochrome camera (Edmund, EO-1312). As shown in Figure 1. The PMF was inserted and embedded in the CRSW, which was used to avoid the polarization disturbance of input light. The CRSW is the solid mixture of $TiO_2$ particles and the OA (Norland, NOA61, $n$=1.53) which acts as a random optical diffusing media in this study. Firstly, the $TiO_2$ particle (average diameter = 5 μm) and the OA were fully stirred by a magnetic stirrer, where the volume ratio of OA and $TiO_2$ particles was ~ 1/1000. The mixed pasty liquid was injected to a mental tube with highly reflective inner wall. Then ultraviolet laser light was incident on the mixture to cure the CRSW. Finally, the UCNPs powder was sprayed on the end face of the CRSW. It is known that the waveguide with larger cross section can provide more speckle information, which is a basis to promote wavemeter's resolution. However, larger cross section will also lower the energy density, which may not be sufficient to excite the UCNPs. After many experiments, the suitable diameter of the CRSW was kept at 1.5 mm.

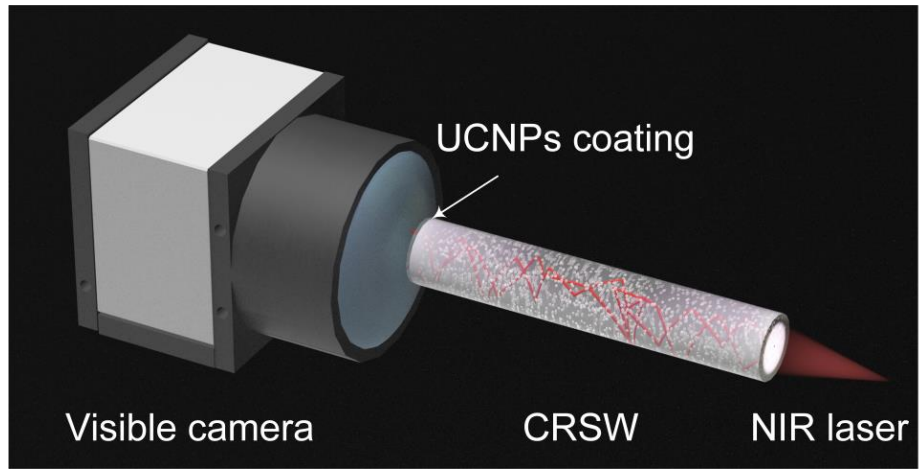

Figure 1 The experimental setup of the designed CRSW wavemeter. Two red lines in the CRSW are typical propagation path of the incident light.

In this study, the UCNPs were prepared by coprecipitation method [23], which account for the light spectrum conversion mainly based on the energy transfer upconversion in Er-sublattice. As shown in Figure 2 (a), these nanoparticles are monodisperse with an average size of ~ 30 nm. The high resolution transmission electron microscopy of the Figure 2 (b) display a clear lattice fringes with a distance of 0.513 nm corresponding to the d-spacing of the (100) lattice planes of hexagonal structure, which reveals the nanoparticle's good crystalline property. The core-shell nanostructure of such UCNPs were confirmed by the scanning transmission electron microscopy (STEM) image and elemental mapping results as shown in Figure 2 (c-f). These UCNPs are in hexagonal phase according to the X-ray diffraction patterns in Figure 2 (g). For the application of proposed wavemeter, the smaller size of UCNPs is beneficial to the resolution of the converted speckle patterns. It is noted that the upconversion emission and absorption spectra of the nanoparticle are critical for the performance of the wavemeter. Figure 2 (h) and (i) shows the absorption spectrum in the range of 900-1700 nm and the upconversion emission spectra under 808 nm, 980 nm, and 1550 nm excitations, respectively. The $NaErF_4@NaYF_4$ UCNPs have an excellent response in the range of 1480-1580 nm, which covers the whole C-band in optical communications.

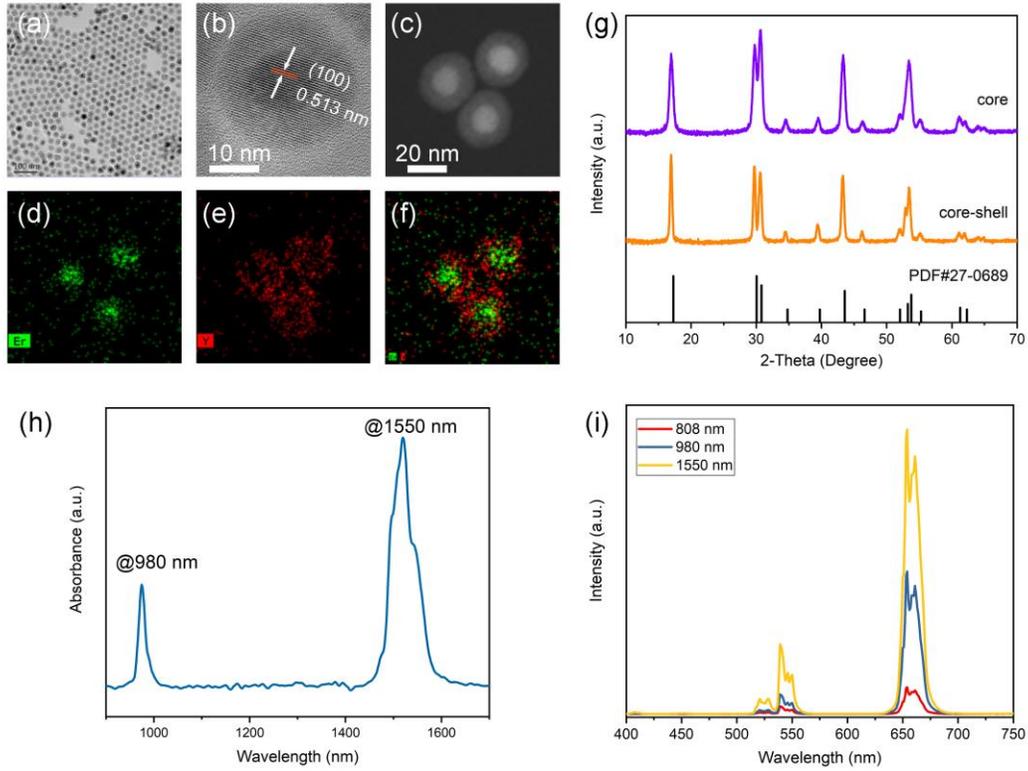

Figure 2 (a) High-resolution transmission electron microscopy (TEM) image and (b) high-resolution transmission electron microscopy of the as-synthesized NaErF$_4$@NaYF$_4$ core-shell UCNPs. (c-f) Scanning transmission electron microscope (STEM) image of the NaErF$_4$@NaYF$_4$ UCNPs and the elemental mappings of Er, Y, and an overlap of them. (g) X-ray diffraction (XRD) patterns of NaErF$_4$ (core) and NaErF$_4$@NaYF$_4$ (core-shell) nanoparticles. (h) The absorption spectrum and (i) upconversion emission spectra under 808, 980, and 1550 nm excitations for the NaErF$_4$@NaYF$_4$ UCNPs.

**Generation of the speckle patterns**

When the light passed through the CRSW, the speckle patterns can be observed at the end face by using a camera. The speckle pattern was arisen from the multiply scattering among TiO$_2$ particles within the CRSW, where the light beams travel along many different random paths. Thus, these wavelength dependent speckle patterns can be regarded as the result of superposition of multi-beam interference. The speckle intensities can be expressed as equation (1) [13, 24, 25]

$$I(l,\lambda) = \sum_{j}^{M}|A_n|^2 + 2\sum_{i \neq j} A_i A_j^* \cos\left[\Delta k_{ij}(\lambda)l\right]$$

(1)

where the first term of the equation (1) represents the non-interference intensity, which varies slowly with the wavelength change. $M$ is the number of light propagation paths, $A_i A_j^* \cos\left[\Delta k_{ij}(\lambda)l\right]$ is interference terms between the path $i^{th}$ and $j^{th}$, the $l$ is the optical propagation length, and $k$ is the wavenumber. The second term of the equation (1) denotes the summation of all different interference paths. From the equation (1), it clearly shows that the intensity distribution of the speckle pattern is strongly wavelength dependent. In addition, according to the equation (1), the maximum phase difference between the longest ($j=M$) and shortest path ($j=1$) can be denoted as $\varphi(\lambda) = \left[k_1(\lambda) - k_M(\lambda)\right] \cdot l$. In order to gain distinct speckles, the phase difference should be approximate to $\pi$ [13]. Obviously, the length of the light propagation path is one of the decisive factors to determine wavelength sensitivity.

**Evaluation of converted speckle patterns**

Four CRSWs with different lengths (3.5, 7, 10, and 13 mm) were manufactured and coated with the same dose of UCNPs in following experiments. Figure 3 (a) ~ (d) show the typical speckle patterns under the same input laser power of 1.5 mW, and Figure 3 (e) ~ (h) show the horizontal intensity distribution of the corresponding typical patterns. The horizontal profiles present the averaged intensities and the fluctuations. More fluctuations mean that the speckle pattern contains richer details. As shown in Figure 3, CRSWs with different lengths have different average intensities. The speckle pattern captured from the CRSW with the length of 3.5 mm has the maximum averaged intensity, but its horizonal profile does not shows a rich fluctuation. While the horizonal profiles with the length of 7 and 10 mm show more fluctuations due to longer propagation length of the incident light. Besides, the speckle pattern of 13 mm long CRSW has the minimum average intensity due to the high insertion loss from absorption and back-scattering. Also, the averaged intensity can reflect the SNR (signal to noise ratio) of the converted speckle images indirectly. Obviously, the converted pattern from the 13 cm CRSW shows a minimum SNR. According to previous studies, lower SNR makes speckle easier affected by ambient light especially when using visible cameras [20, 26]. Therefore, the SNR of the converted images is a non-negligible factor.

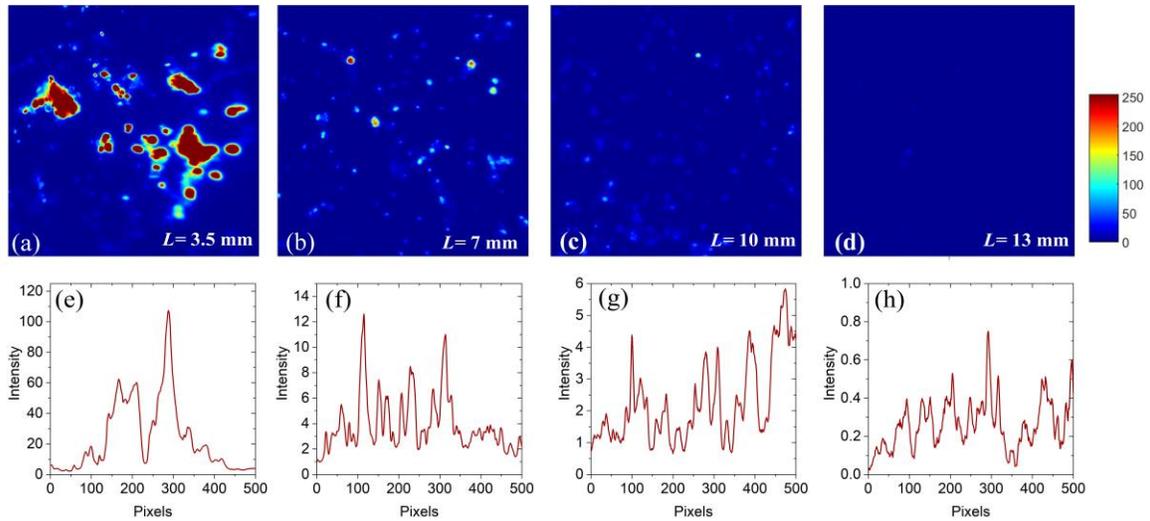

Figure 3 (a-d) The measured speckle patterns of CRSWs with different lengths. (e-h) The horizontal intensity profiles of speckle patterns for (a-d).

The wavelength sensitivity of speckle patterns of four CRSWs were investigated below. The correlation coefficient between patterns plays a decisive role in the resolution of the speckle wavemeter. To intuitively show this, an Arc Cosine Similarity (ACS) algorithm was utilized to express the pattern correlation [20]. This algorithm describes the correlation of two vectors (two speckle patterns in this study) in a geometric perspective. In the following experiments, the wavelength was tunned from 1550 to 1551 nm at a step of 0.01 pm. The calculated ACS values were plotted in Figure 4. It is obviously that the CRSW with the length of 13 mm gains a maximum decorrelation due to its longer light propagation path ($l$) than other CRSWs. With decreasing the length of the CRSW, the correlation of the patterns increase simultaneously. These results further proved that the resolution of the speckle wavemeter closely depends on the CRSW length.

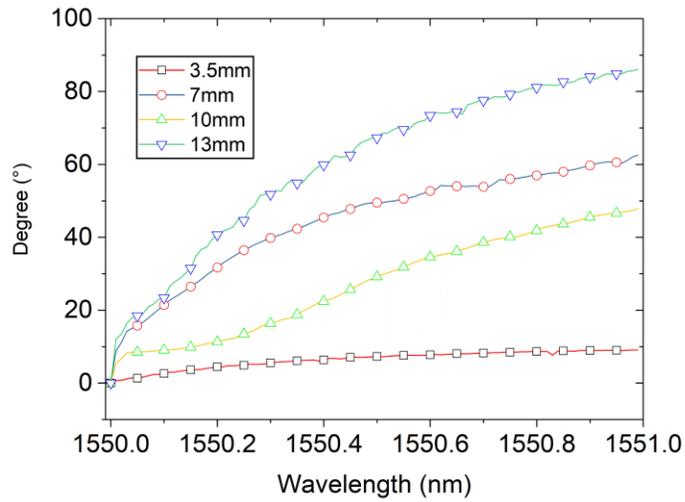

Figure 4 The ACS of speckles with different lengths.

## Results and discussion

### Wavelength reconstruction algorithm

Convolution neural network (CNN) has been widely utilized in optics, such as imaging, sensing and wavelength reconstruction [27, 28]. In this study, a CNN was specially designed to reconstruct the wavelength from the speckle patterns. The network was consisted of 12 layers, as shown in Figure 5. When a speckle pattern was input to the network, it was processed sequentially by a Convolution layer (size=11×11, number of filter = 96, stride = 4×4, and padding = [0 0 0 0]), an activation layer (ReLU), and a max pooling layer (pool size = 3×3, stride = 2×2, and padding = [0 0 0 0]). The convolution layer (Conv) is used to extract the wavelength related features of the speckle patterns. The ReLU layers introduce nonlinear properties to each layer [29]. And the pooling layer is used for down-sampling to highlight the speckles' features [30]. Eventually, the classified wavelengths were obtained through a fully connected layer (FC), a softmax layer (SoftMax), and a classification output layer. The softmax layer denotes the normalized probability of distribution for determining the predicted target. And the $n$ neurons in fully connected layer represent the number of already trained wavelengths [31].

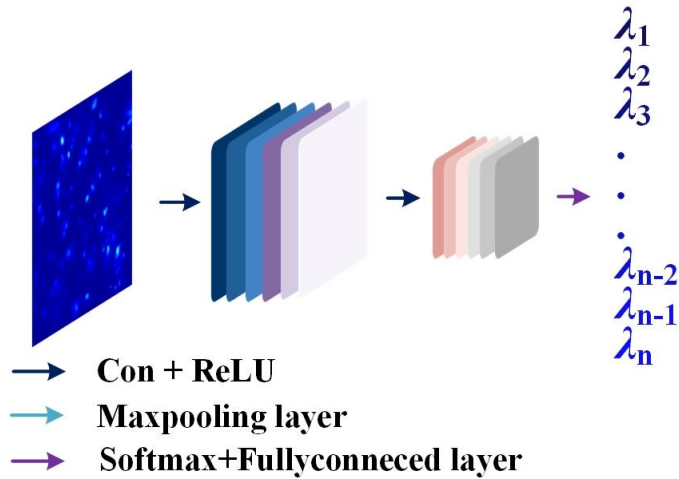

Figure 5 The architecture of the designed CNN. The speckle pattern data is the input of the CNN, and the output is the corresponding wavelength.

In order to test the resolution of the proposed wavemeter, we built up a high precision laser tuning system as the calibration source. The details of this system can be seen in the supplemental materials. In the process of data training, the training set, the test set, and the validation set accounted for 60%, 20%, 20% of the original dataset respectively. And the images of each dataset are randomly sampled from the original dataset. The network training was run on MATLAB through a graphic processing unit (GPU) (Nvidia Geforce GTX 960 4G ) on the platform of Windows. We specified the solver as 'adam' [32]. The learning ratio, the mini-batch size, and maximum number of epochs were set to 0.0001, 48 and 10, respectively. The cross-entropy error function is set as the cost function. [33] The validation result shows that the accuracy of the wavelength achieves to 100%. To further prove the accuracy of the network, we used test image dataset to examine the network. The images of the test dataset were not be trained, and each wavelength has 20 images for examination. The results of wavelength resolution of each group are displayed in the confusion chart as shown in figure 6. The confusion chart is equivalent to a matrix. Each column represents the true wavelength of tested speckle patterns while each row refers to the predicted wavelength. So the diagonal elements correspond to the correctly classified wavelengths and the off-diagonal elements correspond to the incorrectly classified wavelengths. The examination results of each group are listed in Table 1, and it can be seen that the resolution of the wavemeter can be improved by extending the length of the CRSW. The result is consistent with the above calculated ACS. The CRSW with length of 10 and 13 mm has the highest resolution, but the test accuracy of length of 13 mm is not achieve 100%. The main reason is that the light power after CRSW cannot excite the UCNPs efficiently. Due to the low SNR, the network cannot recognize the noise from the ambient light. By contrast, for the cases of 3.5, 7, and 10 mm, their test accuracy are not impacted by the noise level. Therefore, according to the result above, length of 10 mm is the optimal in this study. Generally, the measurement bandwidth of the wavemeter determined by the number of wavelength classes of the original dataset.

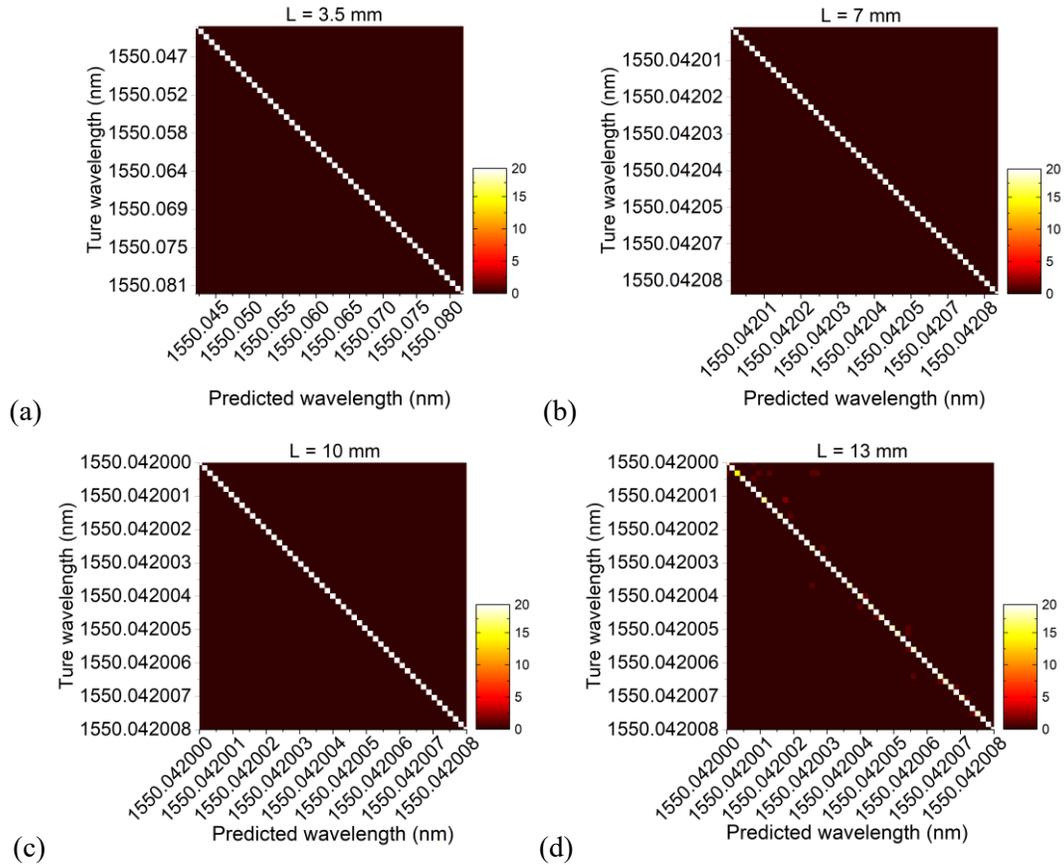

Figure 6 The confusion charts with different lengths of CRSW interrogated by the designed CNN.

Table1 The resolution validation accuracy and wavelength resolution of different length of CRSW.

| Length | Validation accuracy | Wavelength resolution |
|---|---|---|
| **3.5mm** | 100% | 100 MHz (~ 0.8 pm) |
| **7mm** | 100% | 1 MHz (~ 8 fm) |
| **10mm** | 100% | 20 kHz (~ 0.16 fm) |
| **13mm** | 95.3% | ~ 20 kHz (~ 0.16 fm) |

## Transfer learning ability of the trained network

The transfer learning ability of the designed CNN was discussed in the following, which is closely related to the measurement range and the calibration cost of the wavemeter. A technique, t-Distributed Stochastic Neighbor Embedding (t-SNE), was utilized to test the trained CNN according to the ref. [34]. This method is a visible tool that can project the high-dimension data to the low dimension. In this case, it was used to verify whether the untrained speckle pattern is correctly classified by the trained CNN. The speckle patterns around 1530 and 1560 nm were recorded in the verification experiments. For each wavelength, such as 1530 nm, the laser light was tuned by an increment of 0.16 fm for five times (0.16, 0.32, 0.48, 0.64, and 0.80 fm). 100 speckle images were collected for each tuning step. Then these untrained speckle patterns were sent to the trained CNN, and the output of the network's fully connected layers as the input of the t-SNE. As shown in the Figure 7, speckle patterns with different wavelength were clearly clustered. This result shows that the trained CNN has good transfer learning ability and is able to classify the untrained speckle patterns.

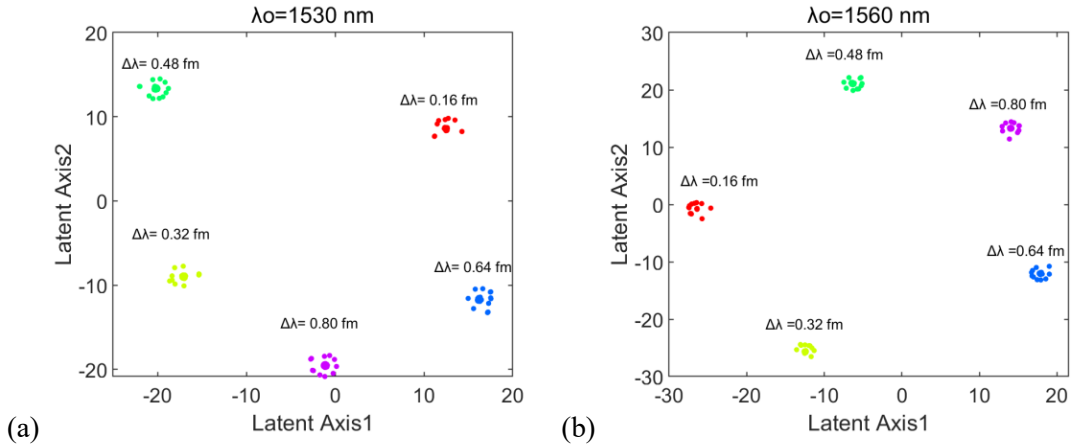

Figure 7 (a, b) The t-SNE visualization of the output of the trained network. The network is trained over the speckle patterns with the wavelength difference of 0.16 fm around (a) 1530 nm and (b) 1560 nm.

## Stability of the speckle wavemeter

The stability of speckles and the associated measurement system is an important indicator, especially for applications. Compared with the MMF in previous studies [20, 26], the CRSW has a more compact size. In the following demonstration experiments, the wavemeter based on CRSW was running without any temperature control. A series of patterns were sampled in the time of

60000 seconds with a time step of 600 seconds at a fixed wavelength of ~ 1550 nm. The ambient temperature change was also recorded, as shown in figure 8. The captured test speckle patterns were then inputted to the trained network, and the wavelength classification results were plotted in bottom row of the Figure 8. It can be seen that the measurement results stay at 1550 nm although there existed fluctuation of temperature and the shot noise of the speckle receiver. The contributions of the wavemeter stability mainly come from the compact volume of CRSW and the robustness of the network. It should be noted that appropriate temperature control setup is necessary to resist larger temperature change.

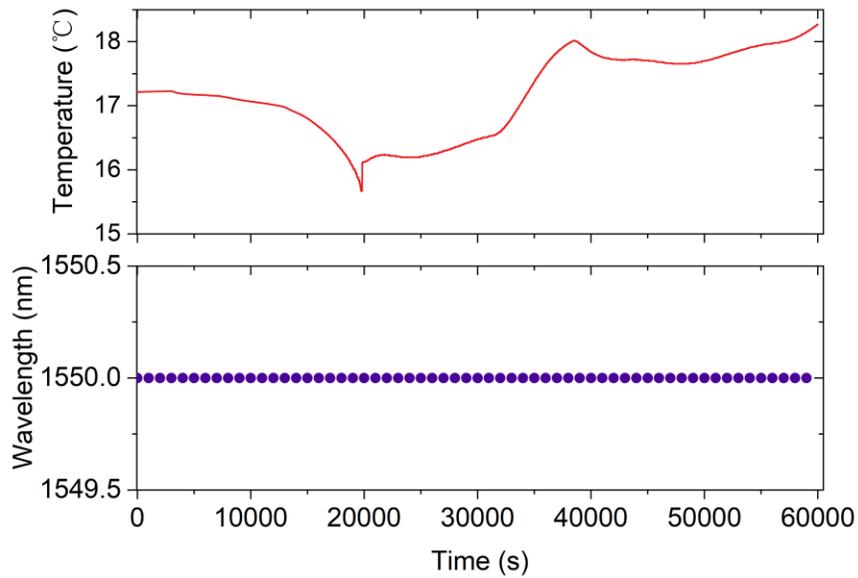

Figure 8 The red curve shows the temperature change within 60000 seconds under the laboratory environment. The blue dot line shows the measured wavelengths every 600 seconds.

## Summary

In this study, we demonstrated that the CRSW made by $TiO_2$ particles and OA, can be used as the optical diffuser in the application of speckle wavemeter. The strong wavelength dependence speckle patterns were arisen from the light multiply scattering between $TiO_2$ particles. The proposed CRSW has the advantages of compactness, stability, low cost and easy making. Furthermore, a thin layer of UCNPs was utilized to convert the NIR to visible light, breaking the response limit of Si cameras in the NIR range. Its special nanostructure ensures high excitation efficient and broader absorption range. Both theoretical analysis and experimental results show that the wavelength sensitivity is closely related to the length of the CRSW. Benefiting from larger optical phase difference, longer CRSWs can gain higher wavelength sensitivity. However, the absorption and back-scattering should not be ignored since the UCNPs need enough light power to excite. Considering the insertion loss, speckle SNR and wavelength sensitivity, 10 mm has been proved to be the optimal length in this study. A CNN was designed for wavelength reconstruction in this study. Based on training hundreds of samples for each wavelength, the CNN can precisely recognize when the speckle change is caused by wavelength or other elements. Compared with other wavelength reconstruction algorithms (one wavelength corresponding to only one calibration sample), CNN enables the wavemeter not severely affected by the shot noise of the camera. The results show that the resolution of the wavemeter achieves 0.16 fm at around wavelength of 1550 nm. Additionally, the t-SNE analysis shows that the trained network also worked well in the untrained wavelengths. Moreover, the results of stability test show that this wavemeter can resist the temperature changes ±2 ℃ without any temperature control due to the compact volume of the CRSW and the robust CNN. In conclusion, this study using a simple CRSW helps generate high wavelength dependent speckle patterns and also pave a new way for the application of UCNPs.

**Acknowledgement**

This work was supported by the National Natural Science Foundation of China (Grant No. 62075202, Grant No. 51972119), the Local Innovative and Research Teams Project of Guangdong Pearl River Talents Program (2017BT01X137), Natural Science Foundation of Zhejiang Province (Grant LY16F050006, Grant LY20F050008), and the Zhejiang Provincial Department of Science and Technology (Zhejiang Xinmiao Talents Program) (Grant 2021R409043).

**Supporting Information**

Supporting Information Available: < The details of the high-resolution laser frequency tuning system.> This material is available free of charge via the Internet at http://pubs.acs.org